# Watching coherent molecular structural dynamics during photoreaction: beyond kinetic description

*Henrik T. Lemke[1,2,*], Kasper Skov Kjær[3,4,5], Robert Hartsock[1,3], Tim Brandt van Driel[4], Matthieu Chollet[1], J. M. Glownia[1], Sanghoon Song[1], Diling Zhu[1], Elisabetta Pace[6], Martin M Nielsen[4], Maurizio Benfatto[6], Kelly J. Gaffney[7], Eric Collet[8], Marco Cammarata[8,*]*

[1]Linac Coherent Light Source, SLAC National Accelerator Laboratory, Menlo Park, CA 94025, USA.
[2]SwissFEL, Paul Scherrer Institut, 5232 Villigen PSI, Switzerland.
[3]PULSE Institute, SLAC National Accelerator Laboratory, Stanford University, Stanford, California 94305, USA.
[4]Molecular Movies, Department of Physics, Technical University of Denmark, DK-2800, Lyngby, Denmark.
[5]Department of Chemical Physics, Lund University, Box 124, Lund SE-22100, Sweden.
[6]Laboratori Nazionali di Frascati-INFN, P.O. Box 13, 00044 Frascati, Italy.
[7]SSRL, SLAC National Accelerator Laboratory, Menlo Park, California 94025, USA.
[8]Institut de Physique de Rennes, Université de Rennes 1, UMR UR1-CNRS 6251, F-35000 Rennes, France.

*to whom correspondence should be addressed: H.T.L., henrik.lemke@psi.ch; M.Ca, marco.cammarata@univ-rennes1.fr

**A deep understanding of molecular photo-transformations is challenging because of the complex interaction between the configurations of electrons and nuclei. An initial optical excitation dissipates energy into electronic and structural degrees of freedom, often in less than one trillionth ($10^{-12}$) of a second. Molecular dynamics induced by photoexcitation have been very successfully studied with femtosecond optical spectroscopies, but electronic and nuclear dynamics are often very difficult to disentangle. X-ray based spectroscopies can reduce the ambiguity between theoretical models and experimental data, but it is only with the recent development of bright ultrafast X-ray sources, that key information during transient molecular processes can be obtained on their intrinsic timescale.**

**In this letter, Free Electron Laser (FEL) radiation is used to measure ultrafast changes in the X-ray Absorption Near Edge Structure (XANES) during the prototypical photoreaction of a spin crossover compound[1,2]. We reveal its transformation from the ligand-located electronic photoexcitation to the structural trapping of the high spin state. The results require a description beyond a kinetic model[3-5] and provide a direct observation of a dynamic breathing of the main structural change. The coherent structural oscillations (period of ~265 fs) in the photoproduct potential lose synchrony within ~330 fs, whereas incoherent motions reveal the energy redistribution and vibrational cooling within ~1.6 ps.**

**We foresee that ultrafast X-ray spectroscopies will provide invaluable insight to understand the complex physics of fundamental light induced phenomena, which are of prime interest in a multitude of chemical[6], physical[7] and biological processes[8,9].**

The prototypical photoactive Spin Cross-Over (SCO) compound [Fe(bpy)$_3$]$^{2+}$ (Figure 1a, where bpy = 2,2'-bipyridine) forms a low spin ground state (LS, S=0) of electronic configuration $t_{2g}^6 e_g^0 L^0$ ($t_{2g}$ and $e_g$ being d-like Fe orbitals and $L$ ligand orbitals)[1,2]. The Fe atom is bound to the bpy ligand by nitrogen atoms, with an average Iron-Nitrogen distance of $r$ ~2 Å.

SCO compounds can be optically excited to an unstable metal-to-ligand charge transfer state (MLCT, $t_{2g}^5 e_g^0 L^1$), which decays to a metastable high spin state (HS, S=2, $t_{2g}^4 e_g^2 L^0$)[1,2]. The ultrafast nature of this process has been investigated by various techniques showing that the MLCT state is short-lived, the HS state is reached on the 100 fs timescale, and coherent vibrations are induced[10-16]. Recent XANES experiments on [Fe(bpy)$_3$]$^{2+}$ helped to clarify the kinetics of the molecular expansion characterized by an r increase of 0.2 Å in the HS structure within less than 200 fs, but without resolving the short-lived MLCT absorption spectrum[3,4]. More recently, X-ray emission experiments have identified an MLCT signature of ~150(50) fs lifetime followed by a short (~70(30) fs) passage through a triplet ($^3$T) state[5].

Here we show time-resolved Fe K-edge XANES measurements of Fe(bpy)$_3^{2+}$ in aqueous solution, after excitation with 530 nm optical pulses (Figure 1b). We used X-ray pulses delivered at the XPP endstation of the Linac Coherent Light Source at the SLAC National Accelerator Laboratory[17-19] (see Methods).

The recorded spectra of the LS ground state as well as 10 ps after photoexcitation (Figure 1c) agree with previous data characterizing the photoinduced change from LS to HS states[3,20]. From the difference signal, we estimate a 75% LS→HS conversion. A comparison with the structurally similar Ru(bpy)$_3^{2+}$ compound[21] indicates that the MLCT spectrum can be approximated by a +1 eV shift of the LS spectrum (MLCT/LS spectra ratio, Figure 1d). High quality transient XANES changes after photoexcitation have been measured at selected energies (Figure 2a) ranging from the pre-edge region (7113.5 eV), with reduced absorption in the HS state as $e_g$ states are occupied[22], up to the 7164 eV, which is mainly sensitive to the molecular structure (small electronic contributions). After an intermediate signal within the first 100 fs at 7121.5 eV and 7132.5 eV, all energies at or above the edge clearly show damped oscillations for $t \gtrsim 200$ fs. At 3-6 ps (inset of Figure 2b and Figure 3), the signals relax to the previously reported HS/LS difference signal[2,20] (Figure 1d).

Considering the similar features in the XANES transients at different energies, we analyze these globally with a phenomenological model including an initially (at t=0) photo-excited state which stochastically populates a final state (*cf.* HS) with a damped oscillatory component (see Methods). The obtained relative amplitudes of that initial state around time zero match the expected MLCT spectrum (Figure 1d). The stochastic transition is therefore interpreted as a decay from MLCT to HS population within $\tau_{MLCT}$=120(10) fs, which activates molecular vibrations ($T_{osc}$=265(10) fs period), damped within $\tau_{osc}$=330(10) fs and accompanied by a slower $\tau_{VC}$=1.6 (0.1) ps component.

The observed 265(10) fs period oscillation (126(3) cm$^{-1}$) in the HS matches the ones observed in optical spectroscopy of Fe(bpy)$_3^{2+}$, which were assigned to non-symmetric Fe-N bending/stretching modes[10,11]. Similar findings from a related SCO crystal discussed the breathing nature of this mode[14,23]. A clear-cut answer on the nature of this vibration - relevant for the SCO mechanism[24] - is of crucial importance for a deeper understanding of the process. Theoretical works on different Fe(II) polypyridine complexes indicate several Fe-N stretching modes of different symmetry in the 110-440 cm$^{-1}$ range[25-27]. The oscillation period observed here is closest to the totally symmetric breathing mode, calculated around 121-125 cm$^{-1}$, corresponding to in phase stretching of Fe-N bonds with relatively rigid bpy. Other modes differing in frequency, symmetry or nature cannot be related to the observed oscillatory signal. The observed XANES oscillations constitute then the direct proof of coherent molecular breathing in the HS potential around $r_{HS}$~2.2 Å.

At energies which are insensitive to the energy shift induced by the change of electronic configuration (so called chemical shift) and for which the change is roughly linear with $r$ (like 7145 eV, Fig. 3; *cf.* XANES signal simulations, Extended Figure 1), we can relate the transient change of amplitude to first order directly to the average structural change of the Fe-N distance. With the amplitude obtained from the global fit (see Extended Table 1), we obtain an initial oscillation amplitude $\Delta r^0 = r - r_{HS} \simeq 0.15$ Å (see Methods). This number, significantly smaller than $r_{HS} - r_{LS} \sim 0.2$ Å, suggests that the HS state is not populated ballistically from the MLCT structure, but by strong incoherent coupling to vibrational modes (Figure 4). $\Delta r^0 \sim 0.15$ Å is very close to the intersection of the calculated $^3T_1$ and HS potentials[26], which might be related to passage through the triplet state observed by X-ray emission[5].

Conversely, the signal at certain energies (like 7156 eV, Figure 3a and Extended Figure 1) has a non-linear dependence with respect to $r$, thus becoming sensitive to higher moments of the wavepacket distribution, *e.g.* its width. Indeed, the experimental data at 7156 eV shows the highest relative contribution of the signal component decaying within 1.6 ps. This narrowing of the distribution $g(r,t)$ in $r$, constitutes a direct observation of the vibrational cooling process in the HS, indicated by optical spectroscopy[10,12,14], but here selectively probed along the Fe-N coordinate. The transient change of the width of a normal distribution in $r$ can be approximately extracted directly from the data (Methods, Figure 3b). The oscillation damping timescale $\tau_{osc} = 330$ fs is attributed to a loss of coherence in the excited wavepacket, rather than a loss of amplitude due to energy transfer (that would otherwise lower the signal at 7156 eV to its value at ~6 ps with a time scale of 330 fs). In order to compare the dataset to the realistically more complex distribution, we built a classical model where stochastically generated "inelastic events" change the phase and the energy of an ensemble of molecular trajectories (see Methods), thereby obtaining a time-dependent distribution of $r$: $g(r,t)$ (Figure 3c). Using the multiple scattering calculations mentioned above, we calculate the expected signals shown in Figure 3b and in Extended Figure 2. The qualitatively superior agreement with the calculation taking into account this temporal distribution $g(r,t)$ over a calculation using its time-dependent average over all distances $\bar{r}(t)$, substantiates XANES' sensitivity to the structural distribution.

Our high resolution data make it possible to go beyond a conventional kinetic description and give a consistent explanation of recent results on photoexcited SCO dynamics, as summarized in Figure 4. We found that the HS potential is reached within 120 fs around $\Delta r \sim 0.15$ Å from its equilibrium structure ($r \sim 2.2$ Å). A HS population within 50 fs, recently proposed based on an observed 50 fs phase shift in ultrafast optical spectroscopy, shows significantly lower agreement with our experimental data (Extended Figure 3). Moreover, the average oscillation of Fe-N distance $r(t)$, obtained by convoluting a directly excited damped oscillatory dynamics $r_1(t)$ with a decaying 120 fs exponential, produces an apparent 50 fs shift consistent with optical and XANES data (Figure 2b). A substantial part of the 2.3 eV photon energy, bringing the system from LS to MLCT state, is dissipated during the first 120 fs required to reach the HS state (Figure 4). Part of the energy is transferred to the HS breathing mode, coherently activated through the coupling between the electronic state and molecular structure. In the potential energy curve picture along the Fe-N distance, the MLCT to HS transition is not horizontal and has a strong vertical component (dissipation of energy to other degrees of freedom), which is very likely the origin of the high quantum efficiency of this process as underlined by theoretical calculations[24].

The present results obtained in a prototypical photoactive molecule illustrate what can be learned with state-of-the-art femtosecond XANES on the photoswitching pathway across different potential energy curves, by disentangling the changes in electronic and nuclear structures. This technique, which can be combined with advanced quantum mechanical calculations[29], will therefore be of great interest to study more complex systems showing light-activated functions and ultrafast energy transfer in chemistry, physics or biology.

## Methods Summary

The time resolved X-ray absorption signal of aqueous solution of $[Fe(bpy)_3]^{2+}$ was measured using the pump/probe technique through total fluorescence as described previously[4] by using a C*(111) double crystal monochromator available at the X-ray Pump Probe station, LCLS[18]. The relative X-ray to optical pulse arrival times were recorded using the timing tool diagnostic[19]. The temporal smearing due to pump/probe group velocity mismatch in the sample liquid was minimized by using a thin (30 µm) round jet. The overall time resolution was found to be ~25(5) fs (RMS) as result of a global fit.

In the fitted model, the signal is composed from a sum of contributions from MLCT intermediate and HS. The HS state signal amplitude is written as a sum of a constant (difference between LS and HS), a damped oscillation (dynamic in HS state), and an exponentially decreasing contribution (vibrational cooling). The fact that the HS originates from the MLCT intermediate is taken into account by numerically convoluting the HS contribution with the exponentially decaying population of the MLCT state. The multiple scattering signals constituting the structural XANES contribution as function of the Fe-N distance $r$ have been simulated using the MXAN program. Those simulations have been combined with a chemical shift of -1.5 eV for the HS state. The MLCT spectrum contribution was accounted for by the +1 eV shifted LS spectrum.

In order to approximate the transient distribution of the Fe-N distance, an ensemble of molecules was classically propagated within the harmonic approximation of the potential energy curves of LS, MLCT and HS state. The transition (MLCT→ HS) was approximated by stochastic events (exponentially distributed with a lifetime of 120 fs). The loss of coherence as well as the vibrational cooling was approximated by stochastically occurring events of velocity reversal and reduction, using empirically found parameters.

## Methods

**Experimental setup**

The absorption signal was monitored through total fluorescence with high time resolution. The setup was very similar to the one described previously[4]; contrary to the previous experiment, the accurate measurement of the relative X-ray to optical pulse arrival times has been recorded using the timing tool diagnostic[19]. The temporal smearing due to pump/probe group velocity mismatch in the sample solution was minimized by using a thin (30 µm) round jet. The sample was excited by 530 nm pulses from Ti:sapphire laser system and an Optical Parametric Amplifier (OPerA Solo, Coherent).

**Global fit of the XANES data at different energies**

Figure 2 shows the time-dependent change of XANES intensity $\Delta I^E(t)$ for different X-ray photon energies $E$, normalized to the XANES intensity in the LS ($I_{off}$) ground state. It is analyzed as a superposition of the signal from the MCLT and HS states: $\Delta I^E(t)/I_{off} = S^E_{MLCT}(t) + S^E_{HS}(t)$

At a given energy $E$, the contribution to the XANES signal of the MLCT state which decays with a time constant $\tau_{MLCT}$ is given by:

$S^E_{MLCT}(t) = A^E_{MLCT} \times P_{MLCT}(t)$

where $A^E_{MLCT}$ is the amplitude of the MLCT, $P_{MLCT}(t) = IRF(t) \otimes [\exp(-t/\tau_{MLCT}) \cdot H(t)]$, $\otimes$ is the convolution operator, $IRF(t)$ is the instrument response function - assumed to be a gaussian $IRF(t) = \frac{1}{\sqrt{2\pi}\sigma} \exp(-t^2/2\sigma^2_{IRF})$ - and $H(t)$ the heaviside function.

The signal from the HS state is described as:

$s_{HS}^E(t) = [A_{HS}^E + A_{OSC}^E \cos(2\pi t/T_{osc} + \phi^E)\exp(-t/\tau_{osc}) + A_{VC}^E \exp(-t/\tau_{VC})] \cdot H(t)$.

where $A_{HS}^E$ is the signal due to the final HS state after cooling, $A_{OSC}^E$ is the starting oscillation amplitude, $\tau_{osc}$ is the damping of the oscillation, $A_{VC}^E$ and $\tau_{VC}$ are the amplitude of the signal and the timescale of the vibrational cooling. Since the MLCT state is the source of molecules in the HS state (the quantum efficiency is close to unity[1,2]), $s_{HS}^E(t)$ is convoluted with $P_{MLCT}(t)$ to give rise to the part of the signal due to the HS state $S_{HS}^E(t) = P_{MLCT}(t) \otimes s_{HS}^E(t)$.

All parameters with an $E$ superscript are allowed to vary independently for each energy except the phases $\phi_E$, kept fixed to 0 or $\pi$. Due to uncertainty in the time zero stability between the different energies, each energy is allowed to be rigidly shifted in time by $t_0^E$. The resulting values of $A_{MLCT}^E$ plotted in figure 1a identify the state as MLCT due to the similarities with the calculated MLCT signal considering as resulting from a +1eV shift.

The physical parameters describing the mechanism with the decay of MLCT to HS accompanied with coherent HS oscillation and vibrational cooling were obtained by a global fitting of the XANES data at different energies in Fig. 2a:

- the time decay of the MLCT $\tau_{MLCT}$=120(10) fs,
- the oscillation period $T_{osc}$=265(10) fs (corresponding to 126(3) cm$^{-1}$),
- the oscillation damping timescale $\tau_{osc}$ =330(10) fs,
- the vibrational cooling decay $\tau_{VC}$=1.6(0.1) ps.
- the RMS width of the IRF ($\sigma_{IRF}$) = 25(5) fs.

The energy dependent parameters along with the uncertainty estimates are reported in Extended Table 1. As the XANES signal at 7145 eV is to first order linear to structural change (Extended Figure 1), its average amplitude can be linearly extrapolated in units of distance from the fit result at that energy: $\Delta r^0 = r_{HS}(A_{OSC}^E/A_{HS}^E)|_{7145\ eV} = 0.15$ Å.

In Figure 2b, the $r_1$ curve is traced using the timescale parameters found by the global fit:
$r_1(t) = r_{HS} - \Delta r^{0,ballistic} [\cos(2\pi t/T_{osc})\exp(-t/\tau_{osc})]$
$r_1(t)$ represents the average trajectory if all molecules would transition in the HS state at t=0 (i.e. if $\tau_{MLCT} = 0$). For visualisation purposes, the amplitude of the full displacement $\Delta r^{0,ballistic}$=0.2 Å was used.

**Estimation of the distribution width from the experimental data**

We start from the EXAFS equation, approximating it as resulting only from the first coordination shell with 6 N atoms at the same distance $r$: $\chi(k,r) = \frac{6f(k)}{kr^2}\exp(-2k^2\sigma^2)\sin(2kr)$, by rewriting $r=r_{HS}+\Delta r$, where $r_{HS}$ is the HS equilibrium distance;

The ratio assuming unchanged thermal factors becomes:
$\Delta I/I = \chi(k,r)/\chi(k,r_{HS}) - 1 = r_{HS}^2/(r_{HS}+\Delta r)^2 \sin[2k(r_{HS}+\Delta r)]/\sin[2kr_{HS}] - 1$.

This can be expanded around $\Delta r = 0$ and for an interference minimum $2kr_{HS} = 3/2 \cdot \pi$ as
$\Delta I/I = -2\Delta r/r_{HS} + \Delta r^2(3/r_{HS}^2 - 2k^2) + \Delta r^3(4k^2/r_{HS} - 4/r_{HS}^3) + \Delta r^4(2/3k^4 - 6k^2/r_{HS}^2 + 5/r_{HS}^4) + O(\Delta r^5)$.

Assuming a normal distribution centered around $\Delta r = 0$, with a width of $\sigma_r$, we obtain
$\Delta I/I = (3/r_{HS}^2 - 2k^2)\sqrt{2\pi}\sigma_r^3 + (2/3k^4 - 6k^2/r_{HS}^2 + 5/r_{HS}^4)\sqrt{2\pi}\sigma_r^5$. For values corresponding to $E = 7156\ eV \rightarrow k = 3.3$ Å$^{-1}$, and $r_{HS} = 2.2$Å, the first term dominates over the second one and we obtain the formula linking the measured signal to the distribution width: $\Delta I/I \approx -44.5\sigma_r^3$.

**Transient molecular distribution model**

To further investigate the origin of the 1.6 ps relaxation, a simple dynamical model for the vibrational mode has been developed which, coupled with XANES calculations, aims to at semi-quantitatively reproduce the observed signal. The idea behind this model is that the loss of phase (dephasing) at minimal loss of energy can be distinguished from damping of vibrational energy (loss of average amplitude) by comparing the transient distribution with the data, which broadens with the first process and narrows with the latter. Moreover, the rather long MLCT lifetime compared with the oscillation period causes a significant loss of phase and helps to explain the relatively small ensemble averaged oscillation amplitude (with respect to the step amplitude).

Briefly, starting with thermally distributed position and momentum of the coordinate $r$, the system evolves in the LS potential (characterized by a vibration period of ~170 fs). Each "molecule" of the ensemble is then "promoted" from the LS to MLCT at a given time (taken from a normal distribution representing the IRF with an RMS width of 25 fs); the stochastic MLCT to HS transition is modeled by an exponential time distribution of 120 fs lifetime ( matching that of the intermediate MLCT). To account for the smaller amplitude of oscillation (0.15 Å) the system is simultaneously shifted by 0.05 Å upon transition from MLCT to HS, thereby approximating structural changes through very short lived intermediate states in a single stochastic process. Once promoted, the system evolves in the HS potential. To account for dephasing and damping, energy redistributions events (average time between them of 400 fs) are considered. They transfer a part of the energy to a thermal bath and cause a phase loss by inverting the momentum. In order to link the dephasing events to the loss of energy, it is assumed that for each event a certain fraction of energy is lost (60% in the model shown). Furthermore, a simple thermostat has been introduced by adding a stochastic change of velocity (normally distributed) to the velocity of the particles. The rms width of this normal distribution is chosen to be $\sqrt{2\gamma K_B T}$ where $\gamma$ represents a coupling parameter, $K_B$ the Boltzmann constant and T the absolute temperature. The value of $\gamma$ has been chosen to keep the distribution width in the LS state to the value found by Daku and Hauser[28]. The calculation is run - with fixed parameters - for a statistical ensemble of $10^6$ "molecules". The program can be obtained by request to the authors. From this ensemble of trajectories, the time dependent distance distribution $g(r,t)$ is calculated. The results of the calculations are shown as 2D false colour plot in Figure 3c. The distribution center of mass moves from $r = 2$ to 2.2 Å for t<0 to several picoseconds, respectively. Within 1 ps the oscillating component disappears and, on longer time scale, the distribution narrows due to cooling in the HS potential. Coupling the obtained distributions with structural XANES simulations (see below) that give the signal as function of energy and displacement ($r$), i.e. $I_{calc}(E,r)$, allows the calculation of the expected signal in two different ways, by integrating the expected signal

$$\Delta I_g = \int I_{calc}(E,r)\, g(r,t)\, dr \text{ or}$$

$$\Delta I_m = I_{calc}(E,\bar{r}) \text{ with } \bar{r}(t) = \int g(r,t)\, r\, dr$$

In other words, while $\Delta I_m$ takes into account only the average position of the reaction coordinate, $\Delta I_g$ uses the full information contained in the distribution of the reaction coordinate. The calculated XANES changes are normalized to XANES intensity in the LS ($I_{off}$).

**XANES signal simulation**

The XAS calculations are made using the MXAN code[30]. It uses the so-called full multiple scattering approach, avoiding any a priori selection of the relevant multiple scattering paths, together with the muffin-tin (MT) approximation for the shape of the atomic potentials[31]. The Coulomb and exchange parts of the total cluster potential were calculated using the total charge density, which was approximated using a superposition of spherically symmetric self-consistent (SCF) atomic charge densities. These atomic charge densities are generated by SCF relativistic Dirac-Fock atomic code that automatically runs in MXAN for all atoms in the Periodic Table. The use of the MT approximation forces the introduction of two free-parameters, the MT radii of the atomic spheres and the interstitial potential that must be optimized in the calculation to get a good agreement between theory and experiment. Such refinement is used to mimic both the SCF potential of the full cluster and compensate for the MT approximation. This is necessary because SCF and non-MT corrections both modify the atomic t-matrices which likewise depend on the muffin-tin radii and the interstitial potential. It points to the possibility of mimicking the effects of the non-MT correction and a full SCF potential by judicious optimization of the radii and the interstitial potential. This theoretical consideration is the basis of the potential optimization procedure normally applied in the MXAN code[32]. Here we have optimized the non-structural parameters by fitting, in terms of non-structural parameters and the Fe-N distance, the low-spin ground-state experimental data coming from LCLS in the energy range from the edge to 60 eV above, obtaining a very good agreement between experimental data and theoretical calculation and a good structural reconstruction with $r$= 2.05±0.04Å. The non-structural parameters derived after the fitting procedure are kept fixed to calculate the series of spectra with a the Fe-N distance ranging from 1.6 Å to 3 Å, with a step of 0.05 Å.

A chemical shift of -1.5 eV has been used for the calculation of the signal for the high spin state.

# Figures

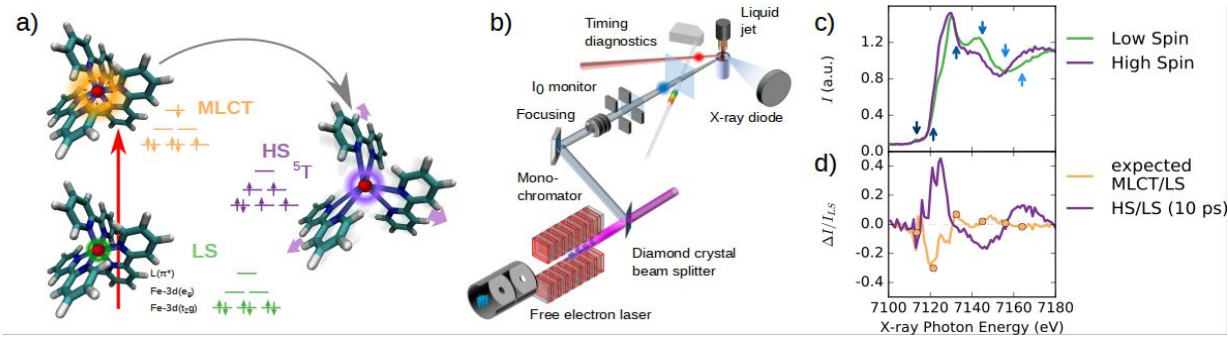

**Figure 1: X-ray Absorption Fingerprints of a Molecular Transformation. a**, Schematics of the photoinduced spin transition from low to high spin for $Fe(bpy)_3^{2+}$. After excitation into an electronic metal-ligand charge-transfer state, the population of antibonding $e_g$ orbitals drives a structural change, characterized by an increase of 0.2 Å of the distance $r$ between the Iron (red) and Nitrogen atoms (blue). **b,** Schematic diagram of the experimental setup. The X-ray beam from a free electron laser is monochromatized by a double diamond (111) crystal and focused to ~10 μm by means of Beryllium X-ray lenses. The $Fe(bpy)_3^{2+}$ was dissolved in water (concentration 30 mM) and circulated via a closed loop system through a 30 μm liquid jet. Such thin sample minimized the effect of temporal broadening due to group velocity mismatch between optical and X-ray beam. **c,** changes between the LS and HS X-ray absorption spectra. Arrows indicate photon energies for which high time resolution data have been measured. **d**, The HS/LS spectra change ratio (magenta line, measured at 10 ps) and the expected ratio between MLCT and LS state (orange solid line). Previous studies on the structurally similar $Ru(bpy)_3^{2+}$ suggested that MLCT excitation shifts the spectrum by ~1 eV, but does not change the spectral shape[21] since the MLCT and LS nuclear structures are similar. The calculated MLCT/LS difference spectrum is approximated by a +1 eV shift of the measured LS spectrum. The dots in panel d) are the measured MLCT amplitude based on a global fit of the data (see text). Both curves have been scaled to 100% excitation.

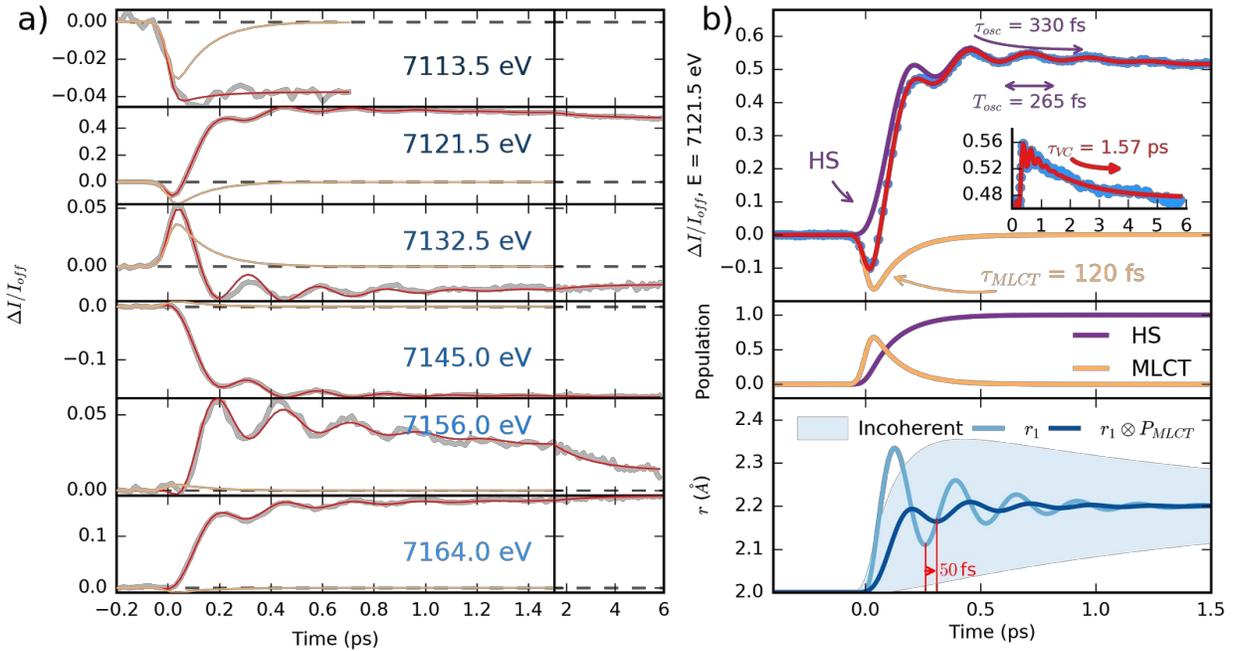

**Figure 2: Following dynamics in real-time: a**, Time scans of relative X-ray absorption change at selected X-ray energies (solid gray lines). For all energies except the pre-edge (7113.5 eV), clear 126(3) cm$^{-1}$ oscillations (265 fs period) are visible. The entire experimental dataset has been globally fitted with the model described in the methods (thin red lines). Essentially, while all amplitude parameters have been varied for every X-ray energy, the physical parameters (MLCT lifetime $\tau_{MLCT}$=120(10) fs, oscillation period $T$=265 (10) fs and damping $\tau_{OSC}$=330 (10) fs, $\tau_{VC}$=1.6(0.1) ps vibrational cooling) are the same for all energies. Orange lines represent the contributions assigned to the MLCT intermediate. **b**, Example fit for 7121.5 eV (top panel), showing the individual contributions of MLCT (orange) and HS (purple) along with the total signal (HS+MLCT, red). The inset shows the data on a longer time window. The model disentangles the kinetic description (MLCT and HS population, mid panel) from the structural dynamics given by the time evolution of $r$ (bottom panel). The exponential grow of the HS population from the MLCT intermediate state leads to an average coherent oscillating trajectory $\bar{r}(t) = r_1(t) \otimes P_{MLCT}(t)$. This has a reduced amplitude and a phase shift by ~50 fs with respect to a directly initiated damped oscillating trajectory $r_1(t)$, which might explain the observations in recent optical data[11]. The incoherent part of the molecular oscillations, the transient distribution width in $r$, decays within 1.6 ps.

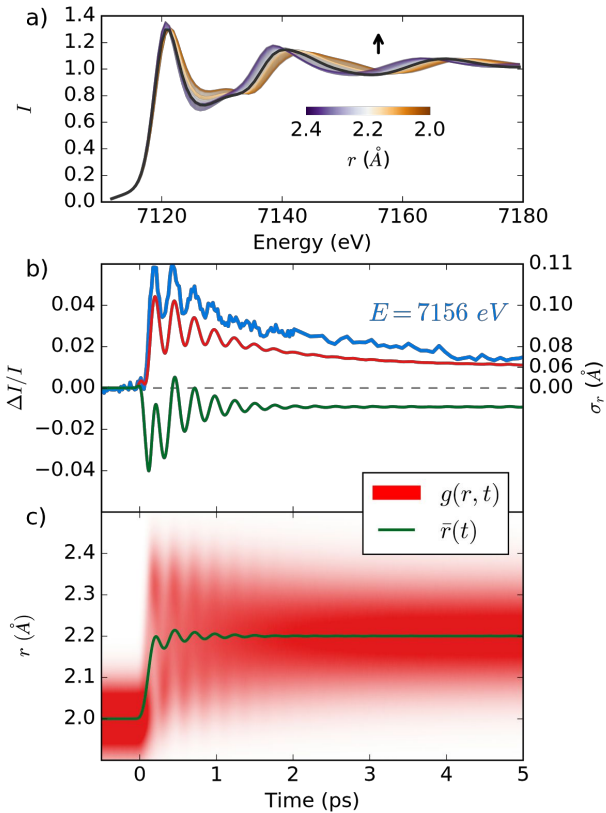

**Figure 3: Coherent vs incoherent structural dynamics: a**, multiple scattering XANES absorption spectra calculated for a variety of distances of $r$ (blue: $r = 2$ Å to red: $= 2.4$ Å) around the average HS configuration ($r_{HS}$ = 2.2 Å). **b**, experimental data at 7156 eV (blue), together with simulated signals calculated using the distance distribution $g(r,t)$ (red) of an ensemble motion model (red in **c**) or the average distance $\bar{r}(t)$ change (green). The right y scale represents the calculated width change $\sigma_r$ of a normal distributed approximation of $g(r,t)$ based on the expansion of the EXAFS equation. **c**, simulated average distance $\bar{r}(t)$ change (green) and distance distribution (red). The data are significantly better reproduced by simulations including the ensemble distribution, consistent with the transient distribution signal $\sigma_r$.

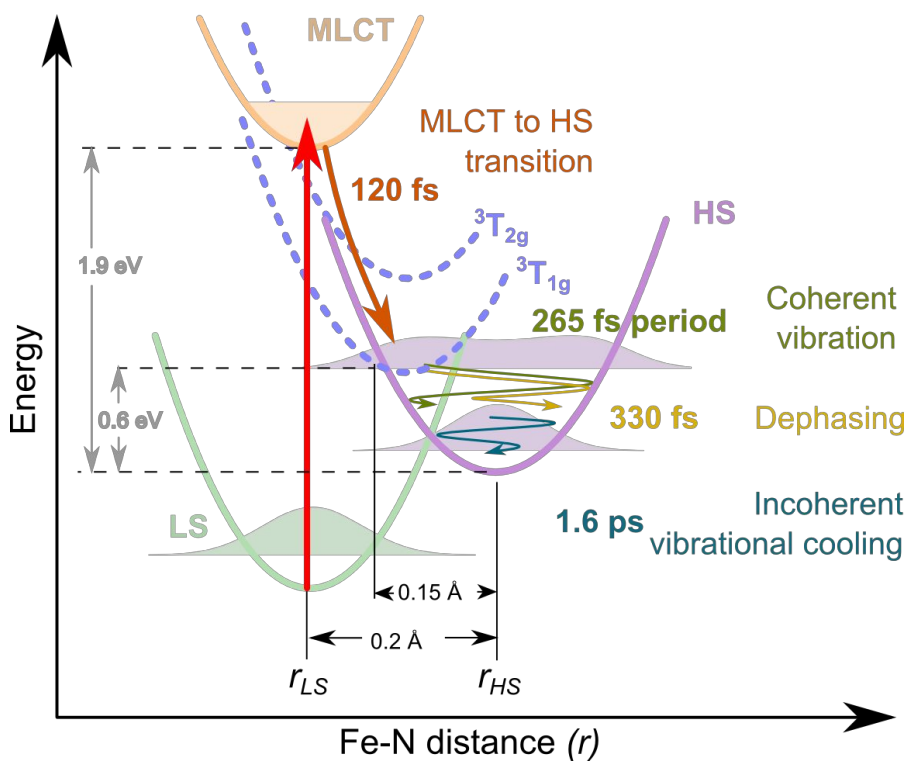

**Figure 4: Intersystem crossing dynamics.** Schematic representation of the observations during the molecular transition in Fe(bpy)$_3^{2+}$ from LS to HS state on the Fe-N distance reaction coordinate. After optical excitation to the MLCT manifold, a decay towards the HS state occurs (120(10) fs) through the triplet state where a large fraction of energy is dissipated to other eigenmodes than breathing. The HS potential is reached around $\Delta r \sim 0.15$ Å from the equilibrium position and the Fe-N distances in the first Iron coordination shell expands and coherently oscillate (265 fs period) around the new equilibrium structure while losing energy. The wave packet disperses at 330 fs time constant and vibrationally cools inside the HS state potential within 1.6 ps (Potential energy curves adapted to calculation results[26]).

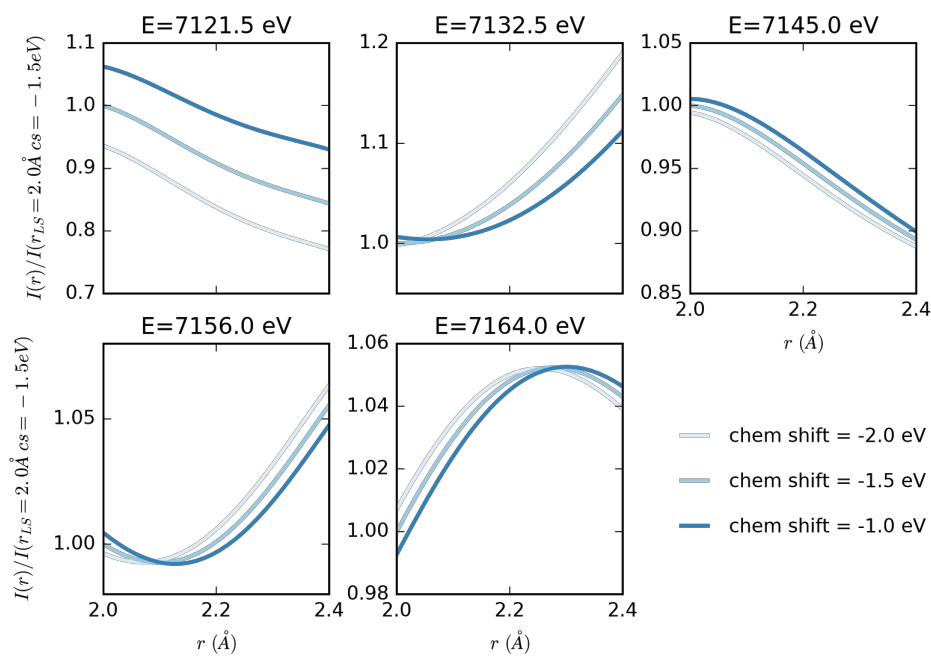

**Extended Figure 1**: Dependence of the signal using the multiple scattering calculations as function of the Iron-Nitrogen distance (*r*) and chemical shift (*cs*). Curves are normalized for the intensity calculated for *r=2 Å* and *cs=-1.5 eV*; At certain energies like 7145 eV the changes are predominantly linear with *r* and show little sensitivity to the chemical shift. Conversely, 7156 eV shows a significant non-linear dependence regardless of the chemical shift.

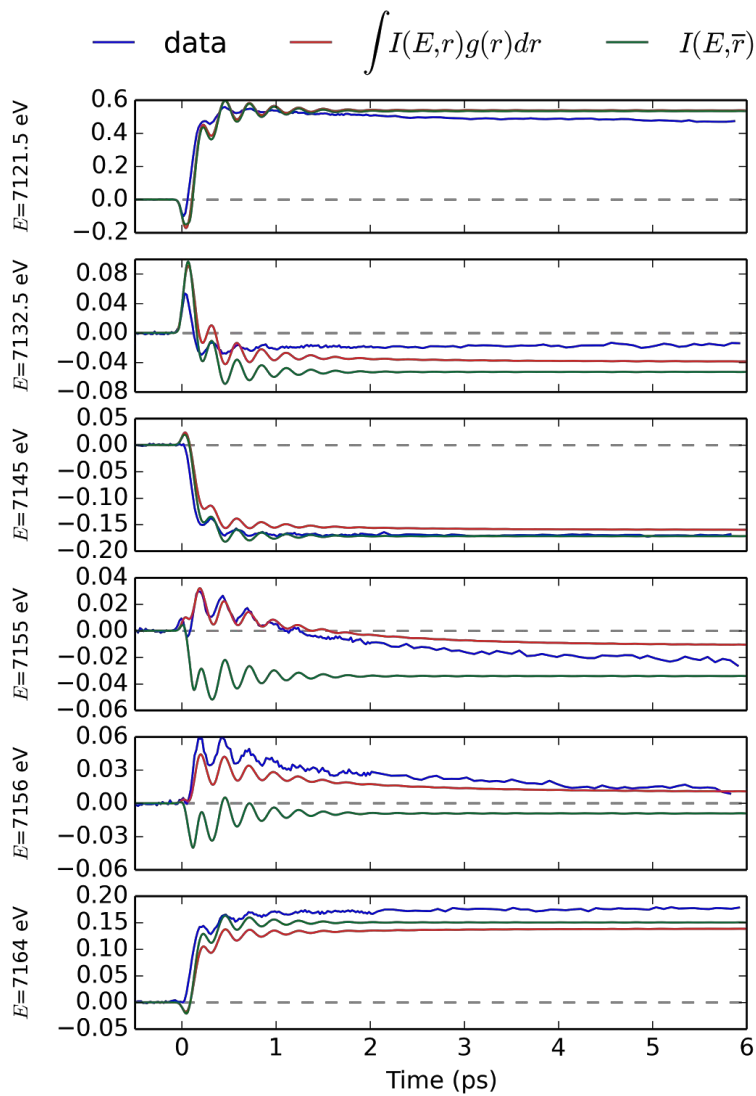

**Extended Figure 2**: Expected signals calculated using the distance distributions *g(r,t)* (red) of an ensemble motion model (red curves) or the average distance $\bar{r}(t)$ change (green curves). The experimental data (blue) is shown for comparison.

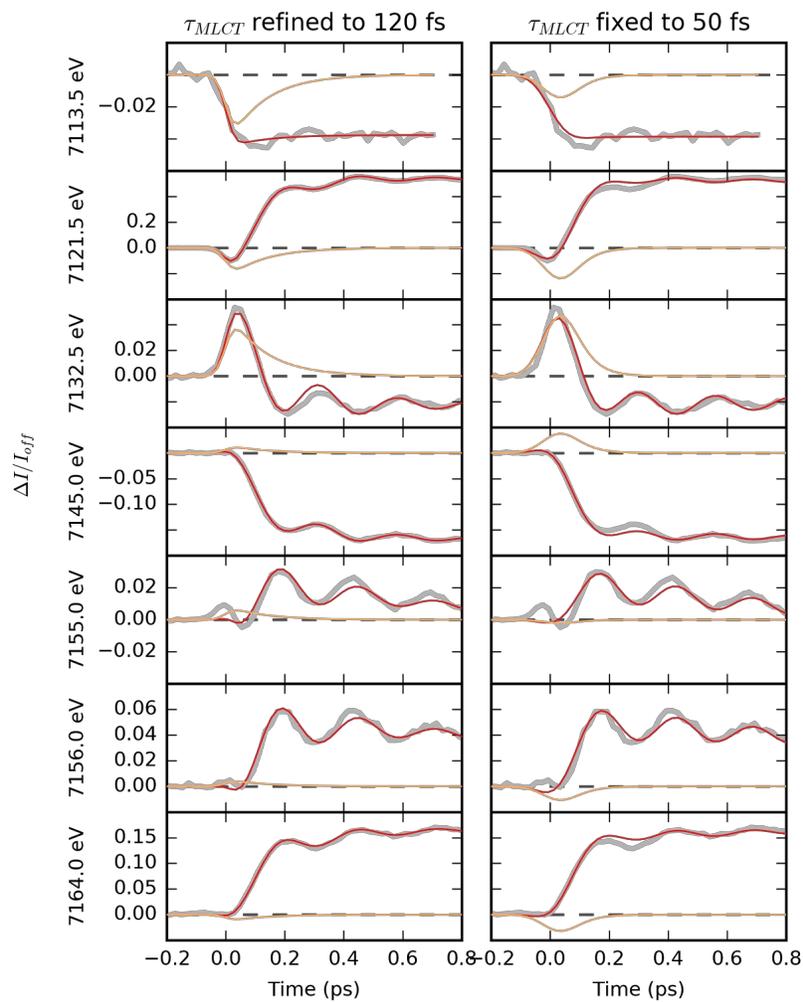

**Extended Figure 3**: Comparison of the global fit model with free (left, same as Figure 2, gray lines: data; red line: fit, orangeline: MLCT contribution) or fixed (right) MLCT lifetime. A clear misfit is present for the 50 fs case especially at short times (0.1-0.3 ps).

| Photon energy (eV) | $t_0$ (ps) | $A^{MLCT}$ | $A^{HS}$ | $A^{VC}$ | $A^{OSC}$ |
|---|---|---|---|---|---|
| 7113.5 | 0.018 ± 0.002 | -0.045 ± 0.001 | -0.038 ± 0.004 | 0.0 (Fixed) | 0.0 (Fixed) |
| 7121.5 | -0.037 ± 0.003 | -0.24 ± 0.03 | 0.476 ± 0.004 | 0.094 ± 0.007 | 0.355 ± 0.005 |
| 7132.5 | -0.01 ± 0.00 | 0.054 ± 0.002 | -0.015 ± 0.001 | -0.01 ± 0.001 | 0.111 ± 0.002 |
| 7145 | 0.003 ± 0.001 | 0.016 ± 0.002 | -0.171 ± 0.002 | 0.005 ± 0.001 | 0.131 ± 0.002 |
| 7155 | 0.005 ± 0.002 | 0.009 ± 0.001 | -0.023 ± 0.002 | 0.048 ± 0.003 | 0.078 ± 0.001 |
| 7156 | 0.006 ± 0.002 | 0.006 ± 0.002 | 0.013 ± 0.002 | 0.041 ± 0.003 | 0.113 ± 0.002 |
| 7164 | -0.008 ± 0.002 | -0.013 ± 0.001 | 0.178 ± 0.001 | -0.021 ± 0.001 | 0.124 ± 0.002 |

**Extended table 1**: Energy dependent parameters obtained from the global fit.

## Acknowledgements
Portions of this research were carried out at the Linac Coherent Light Source (LCLS) at the SLAC National Accelerator Laboratory. LCLS is an Office of Science User Facility operated for the U.S. Department of Energy Office of Science by Stanford University. M.Ca. and E.C. thank ANR (ANR-13-BS04-0002) and Centre National de la Recherche Scientifique (CNRS) (PEPS SASLELX) for financial support. K.S.K, T.B.v.D and M.M.N. acknowledge support from DANSCATT. K.S.K gratefully acknowledges support from the Carlsberg Foundation. K.J.G. acknowledges support from the AMOS program within the Chemical Sciences, Geosciences, and Biosciences Division of the Office of Basic Energy Sciences, Office of Science, U. S. Department of Energy. The authors wish to thanks Dan De Ponte (LCLS) for help with the sample delivery system. H.T.L., E.C. and M.Ca wish to thank Samir Matar (ICMCB, CNRS, Bordeaux) for discussions about the $Fe(bpy)_3$ vibrational modes.

## Authors contributions
H.T.L. and M.Ca. conceived the project and analyzed the data. H.T.L, K.S.K, K.J.G., E.C., M.Ca set the physical picture for interpreting the data. H.T.L, K.S.K, R.H., T.B.v.D, M.Ch., M.J.G., S.S., D.Z, E.C., M.Ca performed the femtosecond XAS experiment. E.P. and M.B. performed the multiple scattering calculations. M.Ca. developed the molecular distribution model and simulated the transient data. H.T.L, E.C., M.Ca wrote the paper. All authors contributed to discussions and gave comments on the manuscript.